\documentclass[prl,twocolumn,showpacs]{revtex4}
\usepackage{graphicx}
\usepackage{amsmath}
\usepackage{color}
\usepackage[normalem]{ulem}
\usepackage{epsfig}

\begin{document}
\title{Stochastic Desertification}
\author{Haim Weissmann  and Nadav M. Shnerb}
\affiliation{Department of Physics, Bar-Ilan University, Ramat-Gan
IL52900, Israel}
\begin{abstract}
The process of desertification is usually modeled as a first order
transition, where a change of an external parameter (e.g.
precipitation) leads to a catastrophic bifurcation followed by an
ecological regime shift. However, vegetation elements like shrubs
and trees undergo a stochastic birth-death process with an absorbing
state; such a process supports a second order continuous transition
with no hysteresis. We present a numerical  study of a minimal model
that supports bistability and catastrophic shift on spatial domain
with demographic noise and an absorbing state.  When the external
parameter varies adiabatically the transition is continuous and the
front velocity renormalizes to zero at the extinction transition.
Below the transition one may identify three modes of
desertification: accumulation of local catastrophes, desert invasion
and global collapse. A catastrophic regime shift occurs as a
dynamical hysteresis, when the pace of environmental variations is
too fast. We present some empirical evidence, suggesting that the
mid-holocene desertification of the Sahara was, indeed, continuous.
\end{abstract}
\maketitle

The catastrophic bifurcation and its statistical mechanics analog,
the first order transition, play a central role in the physical
sciences. In these processes a tiny
 change in the value of an external parameter leads to a sudden jump
 of the system from one phase to another. This change is irreversible
 and is accompanied by hysteresis: once the system relaxed to its new phase,
it will not recover even when the external parameters are restored.

The relevance of these processes to the ecology of population and
communities has been established while ago \cite{may1977thresholds}.
Recently, there is a growing concern about the possible occurrence
of  regime shifts in ecological systems
\cite{scheffer2001catastrophic,rietkerk2004self,kefi2007spatial,scheffer2009early}.
The anthropogenic changes of local and global environmental
parameters  from  habitat fragmentation to the increasing levels of
CO2 in the atmosphere - raise anxiety  about the possibility of an
abrupt and irreversible catastrophe that may be destructive to the
functions and the stability of an ecosystem
\cite{duraiappah2005ecosystems}. This concern
 triggered an intensive search for empirical evidence that may allow one to identify  an impending tipping point,
 where the most popular suggestion is to use the phenomenon of critical slowing
 down
 \cite{dakos2008slowing,scheffer2009early,boettiger2012early,lenton2011early,
dai2012generic,dai2013slower}. Other suggested early warning
indicators, especially for sessile species,  deal with spatial
patterns and the level of aggregation
\cite{scanlon2007positive,kefi2007spatial, sole2007ecology}

Of particular importance  is the process of desertification, which
is considered as an irreversible shift from the "active" vegetation
state to the "inactive" bare soil state, resulting from an increased
pressure (e.g., overgrazing, declines in precipitation). As drylands
cover about $41\%$  of Earth land surface, desertification affects
about 250 million people around the world \cite{reynolds2007global}.
 Various models show that, when the
vegetation state has a \emph{positive feedback}, like an  increased
runoff interception or reduced evaporation close to vegetation
patches, the system supports two attractive fixed points (alternate
steady states) \cite{sole2007ecology,suding2004alternative}. The
bare soil fixed point is stable, since the desert is robust against
small perturbation (a small amount of vegetation) for which the
positive feedback is too weak while the active state is
self-sustained. Accordingly, the system may cross over from
vegetation to bare soil in two routes: First, a disturbance that
pushes the system to the basin of attraction of the bare soil fixed
point, and second, when the vegetation fixed point losses stability,
i.e., when a change of an external parameter takes the system over
its tipping point \cite{rietkerk1996sahelian}.

However, the bare soil  is an \emph{absorbing state}: it corresponds
to a complete destruction of the vegetation (or at least of a given
species), hence it is not affected by noise.   In a finite system
the process must reach eventually  the absorbing state. In the
thermodynamic limit the steady state solution depends on the ratio
between the rate of local extinction and the chance of
recolonization from neighboring sites. This dynamics resembles the
contact (SIS) process \cite{hinrichsen2000non,kessler2012scaling}
where the  extinction transition is continuous and  belongs to the
directed percolation (DP) universality class. Accordingly,  one
should expect a \emph{reversible} second order desertification, with
no jumps, no hysteresis and no tipping point.

Here we present a study of a  minimal model for desertification with
demographic noise. When the external parameter sweep is adiabatic
the transition is indeed DP continuous, as
\cite{kockelkoren2002absorbing} have already pointed out. We analyze
numerically the system beyond the extinction transition point,
showing that it admits different modes of desertification in
different areas of the parameter space. We identify these modes,
consider the effect of the absorbing state on the velocity of the
front, and discuss (qualitatively) the conditions under which the
deterministic first order transition scenario is a reasonable
approximation.

\begin{figure}
\begin{center}
\includegraphics[width=7cm]{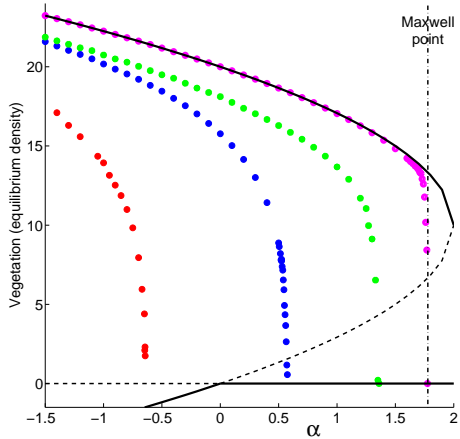}
\vspace{-0.8cm}
\end{center}
\caption{\textbf{The desertification transition}. The lines
represent the possible steady states of the spatially homogenous
solution of Eq. (\ref{GL}) with $\beta=0.4,\gamma=0.02$. Full lines
correspond to stable fixed points, dashed lines to unstable points.
The transcritical bifurcation at $\alpha = 0$ and the saddle-node
bifurcation (tipping point) at $\alpha = 2$ are clearly seen. The
Dash-dot line indicates the  Maxwell point. The symbols are the
steady state density obtained from  numerical solutions of the
process with different $\alpha$s for $D=0.2, \zeta = 30$ (red),
$D=0.2, \zeta = 60$ (blue), $D=10, \zeta = 30$ (green), $D=0.2,
\zeta = 3000$ (purple).  The transition point cannot cross the
Maxwell point.} \label{fig2}
\end{figure}

The model used below is a simple version of the Ginzburg-Landau (GL)
equation where the biomass density $b$ vanishes at one of the fixed
points,
\begin{eqnarray} \label{GL}
\frac{\partial b}{\partial t}=D\nabla^2 b-\alpha b+\beta
b^{2}-\gamma b^{3}.
\end{eqnarray}
Here  $D$ is the diffusion constant, The control parameter $\alpha$
represent the effect of the (changing) environment, $\beta$ is a
positive constant that represents local facilitation, and the
positive constant $\gamma$ accounts for the finite carrying capacity
of the system. When the environment is hostile [positive value of
"stress parameter" $\alpha$, in  Eq. (\ref{GL})] the absorbing
(desert) state $b=0$ is locally stable but local facilitation may
allow the system to have another stable state at a finite vegetation
density. Negative values of $\alpha$ correspond to
 better environmental conditions, where the absorbing state is
unstable (See the bifurcation diagram (lines) in Figure \ref{fig2}).

The deterministic equation (\ref{GL}) admits one or two homogenous
solutions, depending on the value of $\alpha$. Catastrophic
desertification occurs beyond the tipping point, i.e., when $\alpha
\geq \beta^2/(4 \gamma)$, where the system collapses to its desert
state following a saddle-node bifurcation. To recover vegetation,
the stress parameter $\alpha$ should cross zero (transcritical
bifurcation), so  the regime shift is irreversible.

When the initial conditions are inhomogeneous, the desert invades
the vegetation to the right of the Maxwell (melting) point
$\alpha_m> 2 \beta^2/(9 \gamma)$, and vegetation invades on its left
side (see Fig. \ref{fig2}) . The importance of the Maxwell point was
emphasized recently by Bel et. al. \cite{bel2012gradual} (see also
Durrett and Levin \cite{durrett1994importance}): once the system
crosses the Maxwell point, any large disturbance that will generate
a large-enough bare-soil region will invade the rest of the system
and lead to desertification.

Stochasticity in an ecosystem occurs even when rates of demographic
processes (birth, death, migration etc.) are independent of time,
reflecting the randomness of the birth/death process at the
individual level
\cite{scanlon2007positive,kefi2007spatial,manor2008facilitation}.
For example, if $B$ represent a unit of biomass (a shrub, say),  the
quadratic term of Eq \ref{GL} may emerge as the deterministic limit
of the process   $B+B \stackrel{\beta}{\longrightarrow} 3B$, the
cubic term emerges from $B+B+B \stackrel{\gamma}{\longrightarrow}
\oslash$ and the linear term corresponds to  $B
\stackrel{\alpha}{\longrightarrow} \oslash$ (if $\alpha \geq 0$) or
$B \stackrel{\alpha}{\longrightarrow} 2B$ if $\alpha \geq 0$.
\emph{Demographic stochasticity} of this kind yields, for a
population of size $N$, fluctuations amplitude that scale with
$\sqrt{N}$. As mentioned above, in a single site or a finite domain
the system eventually reach  the absorbing state at $b=0$, although
the timescale for this process may be large
\cite{ovaskainen2010stochastic,kessler2007extinction}. For a spatial
system, with migration of individuals to neighboring sites in a rate
proportional to $D$, the system undergoes an extinction transition
when the rate of recolonization of empty sites is equal to the rate
of local extinctions \cite{hinrichsen2000non}.

Eq. (\ref{GL}) appears to be the deterministic limit of this
stochastic process, obtained when $N$, the number of particles per
site (the model is defined off lattice, but any discretization
procedure should involve, at least indirectly, a UV cutoff defined
by the "size" of an individual, or the interaction range), goes to
infinity. This convergence of a stochastic process to the
corresponding PDE's was analyzed  in
\cite{kessler2008novel,kessler2012scaling}, and was shown to fail
close to the extinction transition, when fluctuations govern the
dynamics even in the large $N$ limit. This failure is  limited to a
narrow region close to the transition point, and the width of the
transition zone approaches zero like $N^{-\kappa}$, where the
exponent $\kappa = 2/(d_c-d)$ depends on the upper critical
dimension of the transition and on the dimensionality of the system.
The analysis of \cite{kessler2008novel,kessler2012scaling} takes
into account two prototypes of out-of-equilibrium phase transitions,
the SIS  process \cite{kermack1932contributions} that belongs to the
directed percolation universality class and the SIR process
\cite{kessler2007solution} that belongs to the dynamic percolation
universality class. However, these two transitions are continuous
even in the deterministic limit, while our system admits a
first-order transition when $N \to \infty$

Kockelkoren and Chat\'{e} \cite{kockelkoren2002absorbing} have
already discussed this issue, showing that the  extinction
transition is indeed  second order (DP) once the absorbing state is
taken into account. This result is demonstrated in Figure
\ref{fig2}. Our simulation technique is close to the split-step
method used by \cite{kockelkoren2002absorbing,
moro2004numerical,bonachela2012patchiness}: an Euler integration of
Eq. \ref{GL} (with $\Delta t = 0.001$, $1d$ lattice of $L=10000$
sites, asynchronous update) is  interrupted every $\zeta$
generations when the value of $b_i$ at every site $i$ is replaced by
an integer, taken from a Poisson distribution with an average $b_i$.
We have verified that the transition has, indeed, the critical
exponent of the directed percolation equivalence class
\cite{hinrichsen2000non}.

In Fig \ref{fig2} the equilibrium density is plotted  as a function
of $\alpha$ for different values of $\zeta$ and $D$, together with
the deterministic bifurcation diagram. The stochastic transition is
indeed continuous, but one observes  a new feature: even close to
the deterministic limit (large values of $\zeta$ or $D$) the
transition point cannot cross the Maxwell line. As long as $\zeta$
and $D$  are finite, local extinctions  happen with nonzero
probability \cite{ovaskainen2010stochastic}, and once a local "hole"
is opened, it will spread and overtake the vegetation if the system
is above the Maxwell point \cite{bel2012gradual}. This implies that
the failure of the system to converge to its deterministic behavior
in the $N \to \infty$ limit is not limited to a single point (like
in the SIS/SIR cases) but to a finite domain between the Maxwell
point and the tipping point.

Figure \ref{fig2} also indicates that, when the noise is relatively
weak, the vegetation steady state density decays \emph{linearly} as
the system approaches the extinction transition. The DP theory
predicts a steady state density that scales like
$\Delta^{\tilde{\beta}}$, where $\Delta$ is the distance from the
transition and $\tilde{\beta}<1$ below $d_c$. This is indeed the
case very close to the transition point (result not shown here) but
here the system should converge to the deterministic limit at large
$\Delta$, so the transition region is very narrow and the growth
appears to be linear.

\begin{figure}
\begin{center}
\includegraphics[width=7cm]{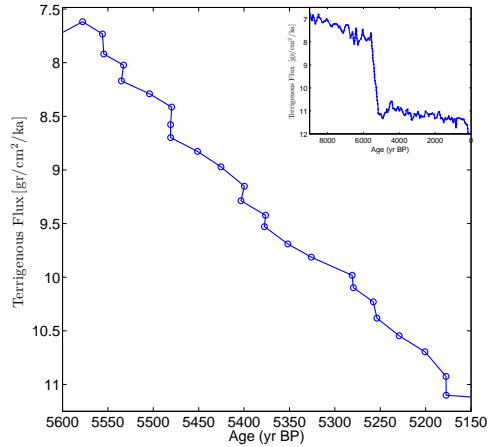}
\end{center}
\vspace{-0.8cm} \caption{The mid-Holocene desertification of the
Sahara, as expressed by the increase of the flux of  terrigenous
dust, during the last 9000 years (inset) and during the transition
period (main panel, modified from \cite{demenocal2000abrupt}). The
transition is assumed to be triggered by a gradual and weak decline
of the Northern Hemisphere summer insolation
\cite{demenocal2000abrupt,scheffer2001catastrophic}.}  \label{fig6}
\end{figure}

The hypothesis of a second-order, reversible desertification
transition with a linear decay of the steady-state density in the
transition regime, is supported by two pieces of data. Reversibility
is suggested by a few recent studies, showing a recovery from
desertification when the external pressure (grazing, in most cases)
has been removed
\cite{fuhlendorf2001herbaceous,rasmussen2001desertification,valone2002timescale,zhang2005grassland,Allington2010973}.
Some evidence for linearity are suggested in Figure \ref{fig6},
where the desertification process of the Sahara during the
mid-Holocene  is traced through the eolian dust record of Site 658C
\cite{demenocal2000abrupt}. The flux of terrigenous sediments seem
to grow linearly during the transition period, in agreement with the
predictions of our model.

Note that the Sahara desertification data are usually interpreted
(see, e.g., \cite{scheffer2001catastrophic})  as an evidence for a
catastrophic, first order transition, since the growth of
terrigenous sediments \emph{percentage} through time appears to be
exponential. However, as stressed in \cite{adkins2006african}, the
use of component percentages in marine sediments can be misleading,
because the total sediment must always sum to 100$\%$. The long
timescales involved (about 500 years) also suggest an alternative
mechanism.

\begin{figure}
\begin{center}
\includegraphics[width=9cm]{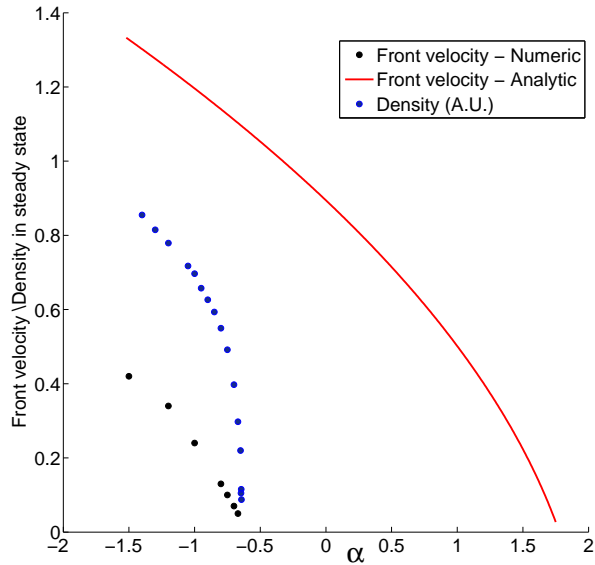}
\end{center}
\vspace{-0.8cm}
 \caption{\textbf{Invasion velocity renormalization.}
Front velocity is shown
 against $\alpha$.  Blue circles represent the steady state density
 of the stochastic simulation for the same value of $\alpha$ (in arbitrary
 units).
 The black dots represent the front velocity  measured in the
 simulation. Parameters  are
 $\beta=0.4,\gamma=0.02, dt=0.01, D=0.2, \zeta=30$.   Given the numerical
 inaccuracies close to the transition,  these two sets
 of circles seem to reach zero at the same point. The red line
 correspond to the analytic expression \ref{eq3}. (for these parameters $\alpha_{MP}=1.778$).
 Front velocity was measured by monitoring the linear growth rate of the $b$ density. The initial conditions
 are vegetation for $5000<x<10000$ and bare soil for $1<x<5000$. }
\label{fig3}
\end{figure}

By studying the system with inhomogeneous initial conditions and
monitoring the growth of the overall density vs. time we have
measured the front velocity $v$. In the deterministic limit the
velocity satisfies,
\begin{eqnarray} \label{eq3}
v=\pm \sqrt{2D}\left(\frac{-\alpha}{m}+\frac{m}{2}\right)
\end{eqnarray}
where
\begin{eqnarray} \label{eq2}
m\equiv
\sqrt{-\alpha+\frac{\beta^2}{\gamma}\left(\frac{1}{2}+\sqrt{\frac{1}{4}-\frac{\gamma
\alpha}{\beta^2}}\right)}.
\end{eqnarray}
The velocity changes sign in the Maxwell point, where the front
changes its characteristic, from a Ginzburg-Landau front to Fisher
type II, at the transcritical bifurcation
\cite{ben1985pattern,kessler1998front}. However, as shown in Figure
\ref{fig3}, under demographic stochasticity the velocity
renormalizes to zero at the extinction transition point.

\begin{figure}
\begin{center}
\includegraphics[width=7cm]{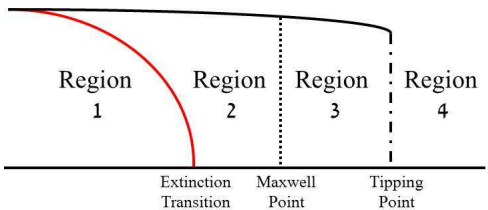}
\end{center}
\vspace{-0.8cm} \caption{\textbf{Modes of desertification - a
schematic . } The steady-state density of vegetation (red line)
approaches zero at the extinction transition. Between this point and
the Maxwell point (region 2) the desertification happens in a series
of local collapses. In region 3 there is also desert invasion, while
in region 4 the collapse is global.   } \label{fig4}
\end{figure}

The emerging insights are summarized in Fig \ref{fig4}. For every
set of parameters (diffusion, noise, nonlinear interaction) the
system admits four different phases.  Above the extinction
transition (region 1) vegetation saturates to an equilibrium value
and will invade a nearby bare-soil region. The steady state density
and the front velocity both vanishes in all other regions, but
desertification takes place in different modes. In region 2 (between
the extinction point and the Maxwell point) the desert does not
invade, and the transition comes about by accumulation of local
extinctions eventuating  a global collapse. In region 3 these
collapses are accompanied by the desert invasion predicted by
\cite{bel2012gradual}
 and the dominant effect
depends on the size of the system and the velocity of the front.
Finally, beyond the tipping point (Region 4) the deterministic
active fixed point loses its stability and vegetation collapse
exponentially, simultaneously all over the place.

All in all, for every system that admits an absorbing state, if
environmental changes (like the rate of variations of $\alpha$) are
adiabatic, the phase transition is a continuous, second order one,
without hysteresis. The catastrophe scenario - a global collapse
after the crossing of the tipping point, followed by an irreversible
transition between alternative stable state, can never be realized.
As long as $\zeta$ is finite, the transition is second order and,
even more importantly, it cannot take place beyond the Maxwell
point, so the tipping point  is completely disparate from the
extinction transition. Accordingly, the attempts to identify an
impending catastrophe by analyzing fluctuation dynamics, utilizing
the critical slowing down as an early warning signal, appears to be
useless.

\begin{figure}
\begin{center}
\includegraphics[width=7cm]{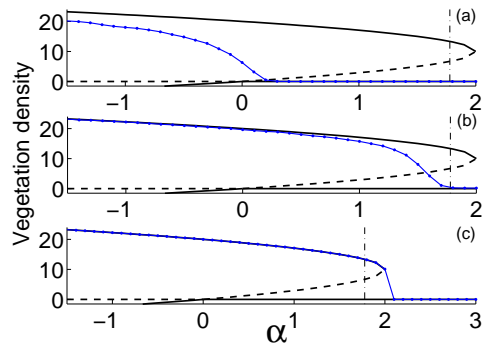}
\end{center}
\vspace{-1.3cm} \caption{\textbf{Dynamical hysteresis}: Vegetation
density (blue) vs. $\alpha$, depicted with the deterministic
bifurcation diagram as a background (black), for $\alpha = -1.5 + s
\cdot t$, $s=10^{-5}$, with $\zeta=40$ (a) $200$ (b) and $1000$
(c).} \label{fig5}
\end{figure}

The studies of catastrophic shifts and early warning signals may be
relevant to the desertification problem only if the environmental
change is non-adiabatic, where the irreversibility has to be
interpreted as a dynamical hysteresis \cite{tredicce2004critical}.
This behavior is demonstrated in Figure \ref{fig5}.  Dynamical
hysteresis is unavoidable close to the extinction transition when
the response of the system becomes slower than the pace of
environmental change, but its effect may be very weak.

As each of the regions 1-4 (in Fig. \ref{fig4}) has its own
characteristic timescale, the conditions for a "rapid" sweep rate
are different in different regions. The deterministic picture is
relevant only when the sweep rate for $\alpha$  is faster than any
other process in the system. However, in such a case the
implementation of critical slowing down indicators close to the
tipping point, assuming that one can trace the relaxation of
fluctuations before the catastrophic shift, may also become
inefficient.

{\bf acknowledgments}  We would like to thank Baruch Meerson, Yoram
Louzoun, Hila Behar,  David Kessler and Ehud Meron for helpful
discussions and useful comments. This work was supported by the
Israeli Ministry of science TASHTIOT program and by the Israeli
Science Foundation BIKURA grant no. 1026/11.

\bibliography{references}

\end{document}